# Game-Theoretic and Reinforcement Learning-Based Cluster Head Selection for Energy-Efficient Wireless Sensor Network

M. Eskandarpour, S. Pirahmadian, P. Soltani, H. Soleimani

Iran University of Science & Technology, Department of Electrical Engineering

*Abstract*— Energy in Wireless Sensor Networks (WSNs) is critical to network lifetime and data delivery. However, the primary impediment to the durability and dependability of these sensor nodes is their short battery life. Currently, power-saving algorithms such as clustering and routing algorithms have improved energy efficiency in standard protocols. This paper proposes a clustering-based routing approach for creating an adaptive, energy-efficient mechanism. Our system employs a multi-step clustering strategy to select dynamic cluster heads (CH) with optimal energy distribution. We use Game Theory (GT) and Reinforcement Learning (RL) to optimize resource utilization. Modeling the network as a multi-agent RL problem using GT principles allows for self-clustering while optimizing sensor lifetime and energy balance. The proposed AI-powered CH-Finding algorithm improves network efficiency by preventing premature energy depletion in specific nodes while also ensuring uniform energy usage across the network. Our solution enables controlled power consumption, resulting in a deterministic network lifetime. This predictability lowers maintenance costs by reducing the need for node replacement. Furthermore, our proposed method prevents sensor nodes from disconnecting from the network by designating the sensor with the highest charge as an intermediary and using single-hop routing. This approach improves the energy efficiency and stability of Wireless Sensor Network (WSN) deployments.

*Index Terms*— Wireless Sensor Network (WSN), Reinforcement Learning (RL), Game Theory (GT), Clustering, Routing.

## I. INTRODUCTION

Wireless Sensor Networks (WSNs) are autonomous systems consisting of spatially distributed sensors that collect environmental data. These sensors transmit acquired information hop-by-hop via wireless communication to a centralized station, where it is processed and control instructions distributed [1]. WSNs are widely used in a variety of applications, including area monitoring [2], healthcare surveillance [3], and earth sensing [4], [5]. Aside from these traditional applications, WSNs are critical for battlefield reconnaissance, critical area surveillance, and emergency response operations. Maintaining continuous sensor operation is critical for a WSN's efficiency and reliability, as it prevents node failures and ensures data flow.

Recent approaches to WSNs have emphasized shortest-path multi-hop routing, which is widely used in clustering algorithms. In these methods, cluster heads (CHs) use routing algorithms to find the shortest path to the sink node, resulting in low delay and a high signal-to-noise ratio (SNR). However, focusing solely on minimizing path length often compromises the long-term stability of the network. Sensors exposed to high data collection rates suffer rapid energy depletion, resulting in uncoordinated node failures. In large-scale networks, battery replacement is expensive and time-consuming, reducing overall network efficiency and coverage.

This paper proposes a hybrid approach that combines clustering and single-hop routing to improve network stability while keeping all sensors operational throughout the network's lifetime. The network is trained using multi-agent reinforcement learning (MARL) to dynamically route data through the most active node. This sensor acts as an intermediary, forwarding information to the sink node. Cluster heads are chosen iteratively using multi-stage clustering, and reinforcement learning ensures that the sensor with the highest battery percentage in each sub-cluster is chosen. Since CH selection is based on real-time energy levels, the designated CHs may change in subsequent iterations, optimizing energy distribution and extending network lifespan.

In recent studies, Frank [23] and Pritsker [24] primarily focused on the stochastic shortest path problem, using multiple integrals to represent probabilistic quantities that needed numerical evaluation for estimation. Due to the rapid increase in computational cost, Mirchandani [25] proposed a reliability computation method to assess link connectivity. However, this approach assumes discrete random variables for edge lengths, imposing a constraint. To address computational inefficiency, Fishman [26], Adlakha [27], and Sigal et al. [28] used Monte Carlo simulation to estimate probabilistic values, but this method is still computationally intensive. Furthermore, these methods require prior knowledge of edge length distributions in stochastic graphs, which limits their usefulness in many real-world scenarios. To address this limitation, Beigy and Meybodi [29] proposed a distributed learning automata (LA) network to find the stochastic shortest path in packet routing. Guo et al. [30] improved the LA-based method by introducing a hierarchical architecture and refining the relevant convergence criteria. Furthermore, Liu and Zhao [31] proposed a whole path feedback mechanism that includes a forced exploration algorithm that uses a random barycentric spanner to explore. Similarly, He et al. [32] used a semi-bandit feedback approach, collecting feedback on individual edges rather than whole paths. Their method strikes a balance between exploration and exploitation by taking distinct paths for each purpose. Additionally, the combinatorial upper confidence bound (CUCB) algorithm [33] and Thompson sampling (TS) algorithm [34] have been used to solve similar problems. Originally designed for multi-armed bandit (MAB) problems, the CUCB algorithm selects multiple simple arms to form a super arm that interacts with its surroundings. Meanwhile, the TS algorithm keeps track of Beta distribution parameters for each action, sampling values from the distribution and choosing the best action based on them. In the context of this problem, various paths can be viewed as actions. Wei et al. [35] developed a novel charging strategy known as CSRL to improve WCE's autonomous path planning capability in WRSNs. This strategy uses Simulated Annealing (SA) to determine actions and traditional reinforcement learning (RL) to find the best charging path for all sensor nodes in the network. However, conventional RL relies on a single agent to learn, which can result in slow convergence. To address this limitation, Iima and Kuroe [36] proposed Swarm Reinforcement Learning (SRL), in which multiple agents learn not only from their own experiences but also by sharing Q-values with one another. SRL has been shown to achieve faster convergence than traditional RL, resulting in more efficient solution discovery. Given these benefits, this study proposes using SRL to improve path planning performance.

The rest of this paper is organized as follows. Section II describes the proposed methodology, while Section III formulates the problem using a multi-agent reinforcement learning framework. Section IV extends this formulation by incorporating game-theoretic principles, and Section V compares the two models. In Section VI, we present a hybrid approach that combines reinforcement learning and game theory, optimizing its implementation to improve the efficiency and longevity of wireless sensor networks

## II. PROPOSED METHOD

a Clustering in a wireless sensor network with a large number of nodes occurs by grouping neighboring sensors together. The clustering is then performed recursively at higher levels among the CHs from the previous phase. While clusters are selected using a predetermined strategy at each stage, CHs are selected independently at each level. Typically, three levels of clustering are sufficient in WSNs, and the process must be designed to ensure that the same number of nodes are present at each step.

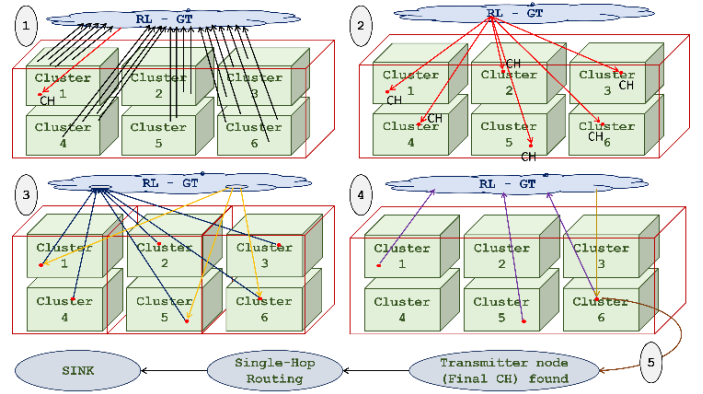

**Fig. 1.** An overview of multi-stage clustering in WSN. In Stage 1, initial clusters are formed among all sensors. In Stage 2, only selected CHs from each cluster are grouped into higher-level clusters. This recursive process continues until a single final CH (transmitter) is identified to send the data to the sink.

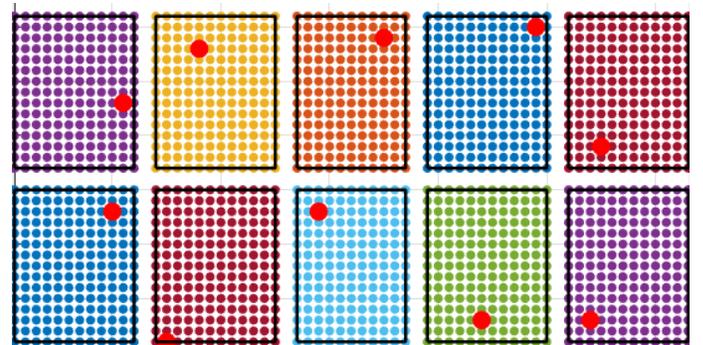

**Fig. 2.** Illustration of selected cluster heads (CHs) in each stage of the clustering process. The CHs from each cluster are re-clustered in the next stage, forming a hierarchy that leads to the selection of the final CH.

In the multi-stage clustering approach, all sensor nodes initially participate in forming the first-level clusters based on proximity. Within each cluster, the node with the highest residual energy is selected as the cluster head (CH). These first-level CHs are then treated as nodes in the next stage, where they undergo clustering again to form second-level clusters. This process repeats recursively for a predefined number of stages (typically three), with CHs being selected at each level based on updated energy levels. Eventually, this leads to a final-stage cluster containing only a few CHs. Among these, the node with the highest remaining energy is selected as the final CH, which acts as a central transmitter, forwarding all collected data directly to the sink node. This hierarchical process ensures balanced energy consumption and avoids overloading any single sensor. To maximize network efficiency, the node with the most residual energy in each cluster is selected as the CH. In the following stage, the process is repeated among the selected CHs, with new clusters created at each level. Eventually, only a few CHs remain, and the transmitter node is chosen during the final clustering phase. Figures 1 and 2 show a comprehensive overview of the Clustering-Based Single-Hop Routing algorithm, which will be discussed in the following sections. Figures 3 and 4 demonstrate that the decision-making process prioritizes selecting the sensor node with the highest charge as the data transmitter, preventing excessive energy depletion in a single node and lowering the risk of critical network failures. This approach avoids scenarios in which a single node repeatedly transmits and receives data, resulting in rapid battery drain. Furthermore, the wireless sensor network's hierarchical clustering structure maximizes network lifetime, balances energy consumption, and prevents premature node failures, thereby improving the WSN's overall stability and reliability. Fig. 3 and 4 further visualize how CHs are selected at each level and how the hierarchy of clusters is formed until the final transmission stage.

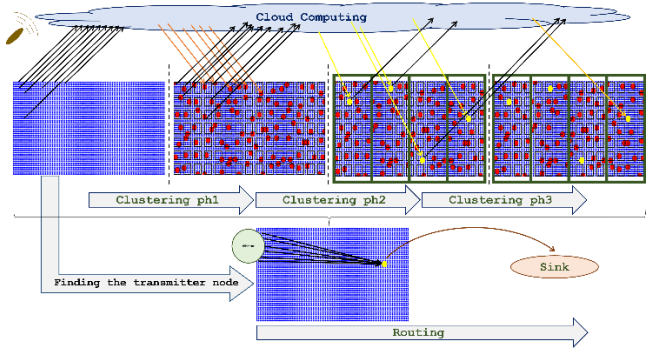

**Fig. 3.** CH selection in each cluster.

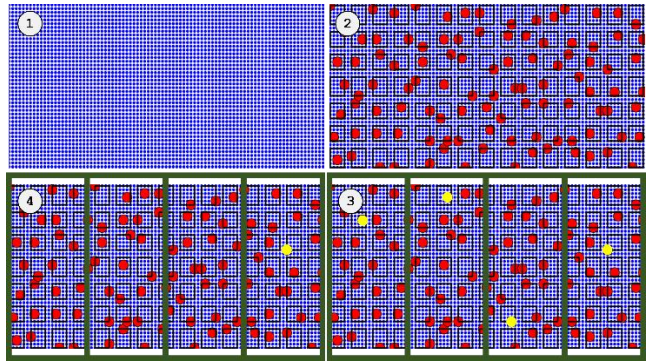

**Fig. 4.** Multi-stage clustering.

## MULTI-AGENT REINFORCEMENT LEARNING

Wireless Sensor Networks (WSNs) consist of many sensor nodes that have limited energy resources, so efficient data transmission and energy management are critical for extending network lifetime. This study introduces a Reinforcement Learning (RL)-based approach to optimizing clustering and routing decisions, resulting in increased energy efficiency and network longevity. Each sensor node is treated as an agent in an RL framework, learning how to form clusters, elect cluster heads (CHs), and send data in an energy-efficient manner. The clustering process continues iteratively until only one cluster head remains to transmit all collected data to the sink node. The RL model makes intelligent decisions using state representation, action selection, rewards, and Q-learning updates, resulting in an optimized hierarchical clustering mechanism.

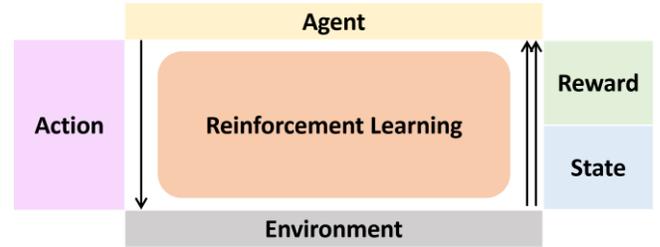

**Fig. 5.** Reinforcement learning Steps.

*States*

In this RL model, the state of each sensor node is determined by energy levels, cluster membership, connectivity, and hierarchical clustering level. These parameters enable the RL agent to evaluate the network's dynamic conditions and make appropriate decisions. To simplify state representation, energy levels are discretized into ten levels (0–9). The cluster membership status indicates whether a node is acting as a cluster head (CH) or as a regular member. Node connectivity is determined by the number of neighboring nodes within communication range. The residual energy ratio compares a node's current energy level to the maximum energy in the network, which influences CH selection. The hierarchical clustering level structures multi-stage clustering, ensuring that clustering continues until a final CH is found. At any given time, each node has a state, which is defined as:

$$S^i_t = (E^i_t, C^i_t, N^i_t, R^i_t, H^i_t) \qquad (1)$$

Where:
$E^i_t$ = Discretized energy level of node $i$ ($E^i_t \in \{0,1,\ldots,9\}$).
$C^i_t$ = Cluster membership status. ($C^i_t \in \{CH, Member\}$)
$N^i_t$ = Number of connected neighbors. ($N^i_t \geq 0$)
$R^i_t$ = Residual energy ratio. $\left(R^i_t = \frac{E^i_t}{E_{max}}\right)$
$H^i_t$ = Current clustering level. ($H^i_t \in \{0,1,\ldots,H_{max}\}$).

*Actions*

At each time step, a sensor node must decide on an action that will affect the clustering and routing process. The available actions are: electing a node as CH, joining an existing cluster, relaying data, re-evaluating clustering decisions, or continuing hierarchical clustering. If a node has enough energy and connectivity, it can elect itself as a CH; otherwise, it joins a nearby CH to stay connected. CHs must also decide whether to send data to a higher-level CH or change their clustering status. For the following steps:
1. Initial clustering of sensors.
2. Cluster head selection based on the highest charge of nodes within each cluster.
3. Secondary clustering of cluster heads.
4. Cluster head selection based on the highest charge of nodes within each cluster head group.
5. Final clustering of cluster heads.
6. Selection of the transmitter node (final cluster head) as an intermediary between the data-receiving sensor and the destination.

The action set is defined as follows:
1. Clustering
2. CH Selecting
3. Single Hopping

*Rewards*

The reward function measures each action's effectiveness in terms of energy efficiency, connectivity, and cluster stability. Positive rewards are provided for using high-energy CHs, maintaining balanced energy distribution, and ensuring high connectivity. Conversely, penalties are imposed for choosing low-energy CHs, increasing energy variance, or causing network disconnections. This ensures that CH selection remains optimal across multiple clustering stages.

Reward Factors:
Valid Clustering Reward: Ensures that no nodes belong to multiple clusters.
o 2 points → If no node is shared between clusters.
o 0 points → If there are shared nodes.

CH Selection Reward: Ensures that the cluster head (CH) is the node with the highest charge in its cluster.
o 3 points → If the CH has the highest charge among cluster members.
o 1 point → If the CH is selected but does not have the highest charge.

Hierarchical Clustering Reward: Ensures that clustering at each stage is performed only among CHs from the previous stage.
o 2 points → If clustering is only among CHs from the previous stage.
o 0 points → If other nodes participate in higher-level clustering.

Final Transmitter Selection Reward: Ensures that the final cluster head (transmitter node) has the highest charge in the entire network.
o 3 points → If the final transmitter node has the highest charge in the network.
o 1 point → If the final transmitter is selected but does not have the highest charge.

Data Forwarding Reward: Ensures that data is correctly forwarded to the final transmitter node for delivery to the destination.
o 2 points → If all data is successfully transferred to the transmitter node.
o 0 points → If data is not transferred to the transmitter.

*Learning Process (Q-Learning Algorithm)*

In reinforcement learning-based clustering for wireless sensor networks (WSNs), the Q-table is an essential component for storing and updating the learned values for each state-action pair. This table enables sensor nodes to make informed decisions about cluster head (CH) selection and routing strategies, resulting in lower energy consumption and a longer network lifetime. The learning process is structured, beginning with the initialization of the Q-table, action selection using an ε-greedy policy, execution of the selected action, observation of the resulting state and reward, and iterative updating of the Q-values using the Temporal Difference (TD) learning principle. Over time, the algorithm improves the clustering strategy by prioritizing actions that result in higher long-term rewards, ensuring peak network performance. The fundamental principle behind this learning method is that each node should learn how to choose the most energy-efficient and stable CH while minimizing energy depletion across the network. The Bellman equation controls the Q-value update in this reinforcement learning system, allowing nodes to evaluate previous decisions and adjust future actions accordingly. This equation is expressed as follows:

$$Q(s,a) \leftarrow (1-\alpha)Q(s,a) + \alpha[r + \gamma \max Q(s',a')] \quad (2)$$

where $Q(s,a)$ represents the Q-value for the current state-action pair, $\alpha$ is the learning rate that determines how quickly the system adapts to new information, $r$ is the immediate reward obtained for taking action $a$ in state $s$ and $\gamma$ is the discount factor that defines how much future rewards influence the current decision. The term $\max Q(s',a')$ represents the highest Q-value that can be achieved from the next state $s'$. This ensures that the system always seeks the best long-term outcome rather than focusing solely on immediate rewards. To effectively explore and exploit different clustering strategies, an ε-greedy policy is used in action selection. The probability ε determines whether a node takes a random action (exploration) or the action with the highest Q-value (exploitation). A high ε value at the start of the learning process encourages the system to explore different options, preventing the algorithm from becoming stuck in suboptimal solutions. As learning progresses, ε decreases to favor exploitation, ensuring that the system applies its acquired knowledge to make more refined decisions. The ε-greedy selection process is mathematically described as follows:

$$a_t = \begin{cases} random\ action & ; \quad probability\ \varepsilon \\ \arg\max Q(s,a) & ; \quad probability\ 1 - \varepsilon \end{cases} \quad (3)$$

where $\arg\max Q(s,a)$ represents the action that yields the highest Q-value for the given state. This mechanism maintains a balance between learning new strategies and applying previously acquired knowledge. After selecting and executing an action, the next step in the learning process is to update the Q-table with the reward received. The reward function is critical in guiding the RL agent toward optimal CH selection. A node receives a positive reward for actions that promote balanced energy distribution, strong connectivity, and long-term network stability. Actions that cause energy imbalances, disconnections, or excessive CH selections are penalized. Experience Replay is an important optimization technique that has been incorporated into the learning algorithm to improve stability and efficiency. Instead of updating the Q-table based solely on the most recent state transition, the system saves previous experiences in a memory buffer and periodically reuses them for training. This technique enables nodes to learn from a wide range of previous decisions, reinforcing positive actions while mitigating the effects of noisy or suboptimal transitions. The Experience Replay mechanism is implemented as follows:

$$Q(s,a) \leftarrow (1-\alpha)Q(s,a) + \alpha[r + \gamma \max Q(s',a')] \quad (4)$$

where previous experiences $(s, a, r, s')$ are sampled from the buffer and used to refine the learning process. This replay mechanism prevents the system from overfitting on recent actions, resulting in a more generalized and robust clustering strategy. To improve convergence speed and efficiency, an adaptive learning rate is used. Instead of a fixed $\alpha$, the system adjusts the learning rate dynamically based on the frequency of state-action visits. The adaptive learning rate is defined as follows:

$$\alpha = \frac{1}{1 + visit(s,a)} \quad (5)$$

where $visit(s, a)$ denotes the number of times the given state-action pair has been encountered. This technique ensures that frequently visited state-action pairs receive smaller updates over time, allowing the system to concentrate on fine-tuning less-explored areas of the state space. The Epsilon Decay Schedule is another important component of the learning algorithm because it controls the rate at which exploration transitions into exploitation. The schedule is expressed as follows:

$$\varepsilon = \varepsilon_0 e^{-\lambda t} \quad (6)$$

where $\varepsilon$ is the initial exploration probability, $\lambda$ is the decay rate, and $t$ represents the current time step. This function ensures that exploration remains high in the early stages of learning while gradually decreasing as the system gains confidence in its clustering decisions. To maintain computational efficiency, the system also uses Q-Table Pruning, which removes rarely used state-action pairs from memory if they do not contribute significantly to decision-making. This technique reduces memory usage and accelerates learning by focusing on the most relevant clustering scenarios.

---

**Algorithm 1:** Reinforcement Learning-Based Multi-Stage Clustering for WSNs

1. **Initialization:**
2.     Define state space $S$ for each node $i$:
3.         $E_i \leftarrow$ Discretized energy level $(0 \dots 9)$
4.         $C_i \leftarrow$ Cluster membership {CH, Member}
5.         $N_i \leftarrow$ Number of connected neighbors
6.         $R_i \leftarrow$ Residual energy ratio $\left(\frac{E_i}{E_{max}}\right)$
7.         $H_i \leftarrow$ Hierarchical clustering level $(0 \dots H_{max})$
8.     Define action space: {Clustering, CH Selection, Data Transmission}
9.     Initialize Q-table $Q(S_i, A)$ with zeros
10.     Set RL parameters:
11.         Learning rate $\leftarrow \alpha$
12.         Discount factor $\leftarrow \gamma$
13.         Exploration rate $\leftarrow \epsilon$
14. **Training Phase:**
15. **for** *each episode* **do**
16.     Initialize network and assign initial states $S_i$ to each node.
17.     **while** *network is operational* **do**
18.         **for** *each node i* **do**
19.             Select action $A_i$ using $\epsilon$-greedy policy:
20.                 Random action if $random(0,1) < \epsilon$
21.                 Else, select $A_i \leftarrow \arg\max Q(S_i, A)$
22.             Execute action $A_i$:
23.                 **Clustering:** Form clusters among neighboring sensors.
24.                 **Cluster Head (CH) Selection:** Choose the node with the highest energy in each cluster.
25.                 **Hierarchical Clustering:** Recursively cluster CHs until a final CH remains.
26.             Compute reward $R_{total}$:
27.                 $+2 \rightarrow$ If clusters have no shared nodes.
28.                 $+3 \rightarrow$ If the chosen CH has the highest charge, else $+1$.
29.                 $+2 \rightarrow$ If clustering occurs only among CHs.
30.                 $+3 \rightarrow$ If the final CH has the highest charge, else $+1$.
31.                 $+2 \rightarrow$ If data is successfully transferred to the final CH.
32.             Update Q-table using:
33. 

$$Q(s,a) \leftarrow (1-\alpha)Q(s,a) + \alpha[r + \gamma \max Q(s',a')]$$

34.             Update state $S_i$ accordingly.
35.     Decrease exploration rate $\epsilon$ over time:
36. 

$$\epsilon \leftarrow \epsilon_0 e^{-\lambda t}$$

37. **End Algorithm**

The Q-learning algorithm was implemented with a learning rate $\alpha = 0.7$, a discount factor $\gamma = 0.9$, and an initial exploration rate $\varepsilon = 1.0$ with exponential decay over time. State transitions and action selections were recorded in a Q-table updated using the Bellman equation. Rewards were calculated based on clustering validity, CH energy levels, and data forwarding success, with values ranging from 0 to 3 depending on the criteria. These parameters were selected after tuning to ensure convergence speed and learning stability across scenarios. Q-learning is particularly suitable for wireless sensor networks due to its low computational and memory complexity. Each node maintains a local Q-table indexed by a discretized state-action space, which allows efficient decision-making without requiring global knowledge of the network. Since sensor nodes operate under tight energy and processing constraints, the use of a simple table-based Q-learning algorithm enables decentralized and scalable operation. The agents learn optimal actions through interactions with their environment over time, enabling them to dynamically adapt to fluctuating energy levels, topology changes, or node failures. This adaptability results in a more balanced energy distribution and prolonged network lifetime, even in large-scale deployments. The localized nature of the algorithm also reduces communication overhead, as decisions are made based on local observations rather than global coordination.

*D. Computational Complexity*

Q-learning, as implemented in our multi-agent WSN environment, operates with relatively low computational complexity, making it viable for sensor nodes with limited resources. The memory complexity is O(|S||A|), where |S| is the number of discrete states and |A| is the number of possible actions. Since we discretize the state space into a manageable number of energy levels, cluster statuses, and connectivity states, the size of the Q-table remains small. For instance, with 10 energy levels, 2 cluster membership states, and a few discrete neighbor counts, the total number of states remains in the order of hundreds, which is acceptable for embedded devices. In terms of time complexity, each Q-learning step involves updating a single Q-value using the Bellman update rule. This operation is O(1), meaning it is constant time per update, and does not depend on network size. The action selection uses an ε-greedy strategy, which only requires a simple comparison over the small action set, also making it O(1). Overall, the per-node complexity of each iteration remains constant and efficient. Moreover, experience replay and adaptive learning rate scheduling add minimal overhead while significantly improving learning stability and convergence. In practical deployments, these components can be selectively included or omitted depending on hardware limitations, making the system flexible. This design ensures that our RL-based method can scale to large WSNs while remaining computationally lightweight and adaptive.

GAME THEORY-BASED CLUSTERING APPROACH

In this section, we present a game-theory-based clustering approach for wireless sensor networks (WSNs) that optimizes energy consumption while extending network lifetime. The goal is to cluster sensor nodes iteratively, selecting the cluster head (CH) based on energy availability, and eventually determining a final CH that acts as the data relay to the sink node. The method ensures uniform energy depletion across sensors, preventing nodes from disconnecting prematurely from the network. We define a set of NN sensors, denoted as $S = \{s_1, s_2, ..., s_N\}$, each characterized by an initial energy level $E_i$ and a communication range $r_i$. Each sensor communicates with its neighbors and chooses to participate in clustering based on its utility function. The clustering process is hierarchical, with sensors forming clusters, electing a CH, and then performing additional clustering among CHs until only one CH remains.

Each sensor $s_i \in S$ formulates a strategy set consisting of two primary choices: (1) become a cluster head or (2) join an existing cluster. The selection of CHs is modeled as a non-cooperative game between sensors, with each node aiming to maximize its utility function. The utility function $U_i$ for a sensor $s_i$ is defined as follows:

$$U_i = \alpha E_i - \beta d_i - \gamma N_i \qquad (6)$$

Where $E_i$ is the residual energy of sensor $s_i$, ensuring CHs have sufficient energy for prolonged operation, $d_i$ is the average distance from $s_i$ to other sensors in its communication range, minimizing transmission costs, and $N_i$ is the number of neighboring sensors that would join the cluster, ensuring optimal load balancing and $\alpha, \beta, and\ \gamma$ are weighting factors that adjust the significance of each parameter. Each sensor independently evaluates its utility and chooses the strategy that maximizes its expected gain. The CH is chosen using an iterative best-response dynamic, ensuring that at equilibrium, no sensor benefits from unilateral deviation. CHs are the sensors that provide the most utility in each local neighborhood. Once the first level of clustering is complete, the CHs form a new network and perform another round of clustering. The process is iterative, with each CH network selecting higher-level CHs until only one remains. This final CH becomes the primary relay node, collecting data from all remaining sensors and sending it to the sink node. The routing strategy uses a single-hop model, in which sensors within a cluster send data directly to their CH. This approach reduces communication overhead and increases the network's operational lifetime. Once the final CH has been determined, all sensors, including previous CHs, send their data to the ultimate CH, which then sends the aggregated information to the sink. To ensure fairness and energy efficiency, the process is dynamically repeated after each clustering cycle using updated energy levels.

**Algorithm 2:** Game-Theory-Based Multi-Stage Clustering for WSNs

1. **Initialization:**
2.    Define the set of sensor nodes: $S \leftarrow \{s_1, s_2, ..., s_N\}$
3.    Each sensor formulates a strategy:
4.       **Become CH**
5.       **Join a cluster**
6.    Assign each sensor an initial energy $E_i$ and a communication range $r_i$.
7.    Define the utility function:
8. 
$$U_i \leftarrow \alpha E_i - \beta d_i - \gamma N_i$$

   where:
9.    $E_i \rightarrow$ Residual energy of sensor $s_i$
10.    $d_i \rightarrow$ Average distance from $s_i$ to other sensors within range
11.    $N_i \rightarrow$ Number of neighboring sensors that would join the cluster
12.    $\alpha, \beta, \gamma \rightarrow$ Weighting factors adjusting parameter significance
13. **Cluster Formation:**
14. **for** each sensor $s_i \in S$ **do**
15.    Compute utility $U_i$.
16.    Choose the strategy that maximizes $U_i$.
17.    Sensors form clusters accordingly.
18. **Multi-Stage Clustering:**
19. **while** number of CHs $> 1$ **do**
20.    **for** each CH **do**
21.       Perform higher-level clustering among CHs.
22.       Recompute utility function for CHs.
23.       Select new CHs at each stage until only one remains.
24. **Final CH Selection and Data Transmission:**
25. The final CH transmits aggregated data to the sink node.
26. **End Algorithm**

Nodes with depleted energy are replaced as CHs by other sensors with more residual energy, ensuring network stability and consistent energy consumption. The iterative clustering mechanism ensures that energy dissipation across the network is balanced, preventing specific nodes from failing prematurely due to high energy consumption. In conclusion, this game-theory-based clustering algorithm provides an efficient and adaptive method for organizing WSNs. It achieves optimal energy usage, load balancing, and network lifetime by utilizing utility-based CH selection and hierarchical clustering. The approach effectively prevents sensor nodes from disconnecting prematurely due to uneven energy consumption, ensuring consistent data transmission to the sink node. This framework can be expanded to include mobility models, adaptive learning strategies, and multi-hop communication for even greater efficiency in real-world deployments.

## HYBRID ALGORITHM

In this section, we present two hybrid approaches that integrate Reinforcement Learning (RL) and Game Theory (GT), each contributing complementary strengths. While GT offers structured, utility-based clustering with fast convergence and decentralized decision-making, it lacks adaptability to dynamic energy conditions. On the other hand, RL enables agents to learn from past decisions, adaptively optimizing CH selection for balanced energy use. By combining them, we leverage the strategic formation of clusters from GT with the adaptive, reward-driven CH selection from RL, or vice versa. These combinations allow the system to maintain both network stability and long-term performance, outperforming pure RL or pure GT approaches alone. The following two approaches—GT+RL and RL+GT—demonstrate how the synergy between both protocols leads to improved energy efficiency and extended network lifetime.

*Approach 1: Clustering with Game Theory, CH Selection with Reinforcement Learning*

In this approach, clustering is performed using game theory, with each sensor evaluating a utility function to determine its membership. To optimize long-term energy efficiency, CH is selected within each cluster using reinforcement learning. Each sensor formulates a strategy set:

1. Become a CH.
2. Join a cluster.

Each cluster uses RL to select the CH. The RL agent observes states including:

- Residual energy.
- Number of cluster members.
- Distance to the cluster centroid.

The action space consists of:

- Assigning itself as CH.
- Joining a neighboring CH.

The process repeats until only one CH remains, transmitting data to the sink as is mentioned in algorithm 3.

**Algorithm 3:** Clustering with Game Theory, CH Selection with Reinforcement Learning

1. **Initialization:**
2.    Each sensor evaluates its utility function (Formula 9) to decide strategy:
3.       **Become a CH**
4.       **Join a cluster**
   Sensors form clusters based on game theory.
6. **CH Selection with Reinforcement Learning:**
7. **for** each cluster **do**
8.    RL agent observes states and selects CH.
9.    Reward function ensures energy-efficient CH selection (Formula 3).
10. **End Algorithm**

*Approach 2: Clustering with Reinforcement Learning, CH Selection with Game Theory*

This method uses RL to cluster data, and game theory to determine CH selection within each cluster. Each sensor learns to cluster dynamically based on energy, connectivity, and spatial distribution.

State Space:

- Residual energy.
- Number of neighboring sensors.
- Distance to potential CHs.

Action Space:
- Form a new cluster.
- Join an existing cluster.

Once clusters have been formed, game theory is used to select CHs. The CH was chosen as the sensor with the highest utility value from algorithm 4.

**Algorithm 4:** Clustering with Reinforcement Learning, CH Selection with Game Theory
1. **Initialization:**
2.     Each sensor learns to form clusters dynamically.
3.     Sensors define state space:
4.         Residual energy
5.         Number of neighboring sensors
6.         Distance to potential CHs
7. **Clustering with Reinforcement Learning:**
8. **for** each sensor $s_i \in S$ **do**
9.     Observe states and form clusters based on RL.
10.     Apply reward function (Formula 3).
11. **CH Selection with Game Theory:**
12. **for** each cluster **do**
13.     Compute utility $U_i$ and select CH based on game theory.
14. **End Algorithm**

### PERFORMANCE EVALUATION

To evaluate the effectiveness of these two hybrid clustering approaches, we ran simulations that measured energy consumption, load balancing, and network lifetime. One of the key performance indicators is the network's average energy level over time. The first method, in which clustering is performed using game theory and CH selection is optimized using reinforcement learning, showed superior energy preservation. This method ensures that CHs are dynamically selected in an energy-efficient manner, resulting in a more gradual and balanced energy depletion across the system. In contrast, the second method, which relies on reinforcement learning for clustering and game theory for CH selection, showed a slightly faster energy consumption rate due to the initial exploration phase of RL, where some suboptimal clustering decisions were made. Another significant metric is the variation in sensor energy levels over time.

TABLE I
SIMULATION PARAMETERS AND COMPARISON

| Parameters | Our Method | Normal Routing |
|---|---|---|
| Area Size | $100 \times 100$ | |
| Number of Nodes | 100 | |
| Packet Size | 4000 bits | |
| Coverage | $r = 0.5$ | |
| Avg SoC (%) after $\frac{T}{10}$ | 95 | 95 |
| Avg SoC (%) after $\frac{3T}{10}$ | 77 | 78 |
| Avg SoC (%) after $\frac{5T}{10}$ | 51 | 53 |
| Avg SoC (%) after $\frac{7T}{10}$ | 36 | 39 |
| Avg SoC (%) after $\frac{9T}{10}$ | 16 | 21 |
| Eliminated Nodes | **0** | 59 |
| Network Longevity | **100%** | 41% |

A lower variance indicates uniform energy depletion, which prevents premature sensor failures. The first approach maintained a more consistent energy distribution because the CH selection process using RL adjusts over time to optimize energy balance. However, the second approach had a higher variance, implying that some sensors depleted energy more quickly, potentially leading to early network fragmentation. Figures 6–9 compare the hybrid methods of game theory and reinforcement learning, highlighting their advantages in various aspects. Figures 10-13 depict the clustering process and final node selection for single-hop routing using full RL, full GT, GT+RL, and RL+GT algorithms, as well as how the overall scenario unfolds. Table 1 provides a detailed comparison of these four algorithmic approaches by evaluating factors such as final received reward, learning algorithm convergence, battery variance over time, and average network charge.

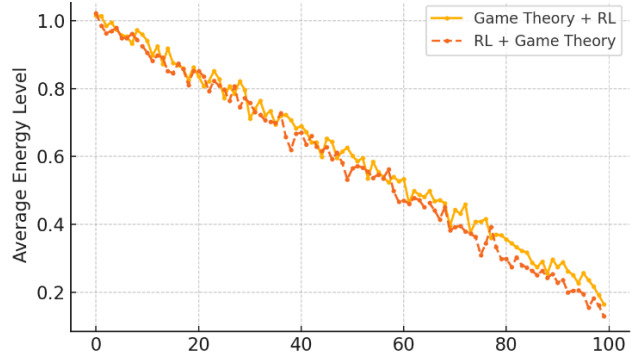
**Fig. 6.** Average energy level of the sensors over time.

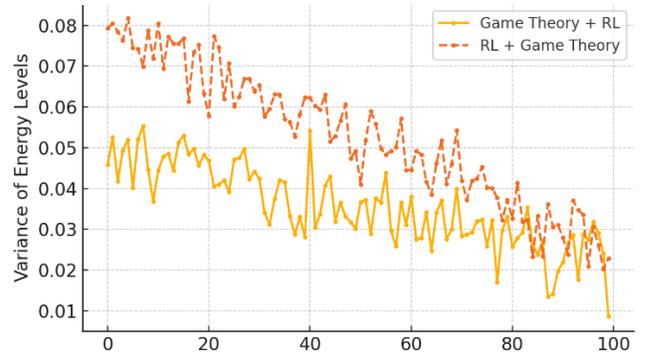
**Fig. 7.** Variance of energy level of the sensors over time.

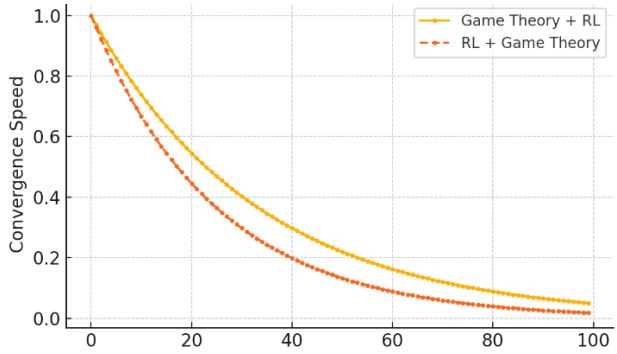

**Fig. 8.** Convergence speed of the algorithms over time.

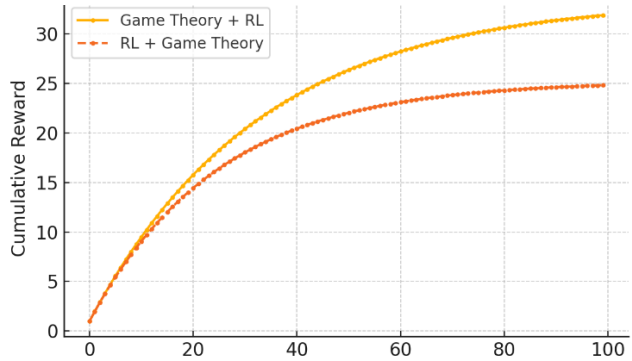

**Fig. 9.** Rewards earned over time.

In addition to energy efficiency and network lifetime, we measured the average end-to-end delay, defined as the time taken for a data packet to travel from a source sensor to the sink via the selected cluster heads. Delay was computed for each approach by calculating the average number of hops and processing steps during each transmission cycle. Results showed that Full RL and GT+RL approaches exhibited slightly higher delays during initial training due to exploration steps in RL. However, as the models stabilized, delay values decreased and remained consistent. Full GT, while quicker in CH selection due to its deterministic nature, led to unbalanced energy usage and earlier node failures, which eventually increased delay due to route breaks and re-clustering needs. On average, GT+RL had the most balanced delay profile, maintaining both stability and low routing depth due to effective clustering.

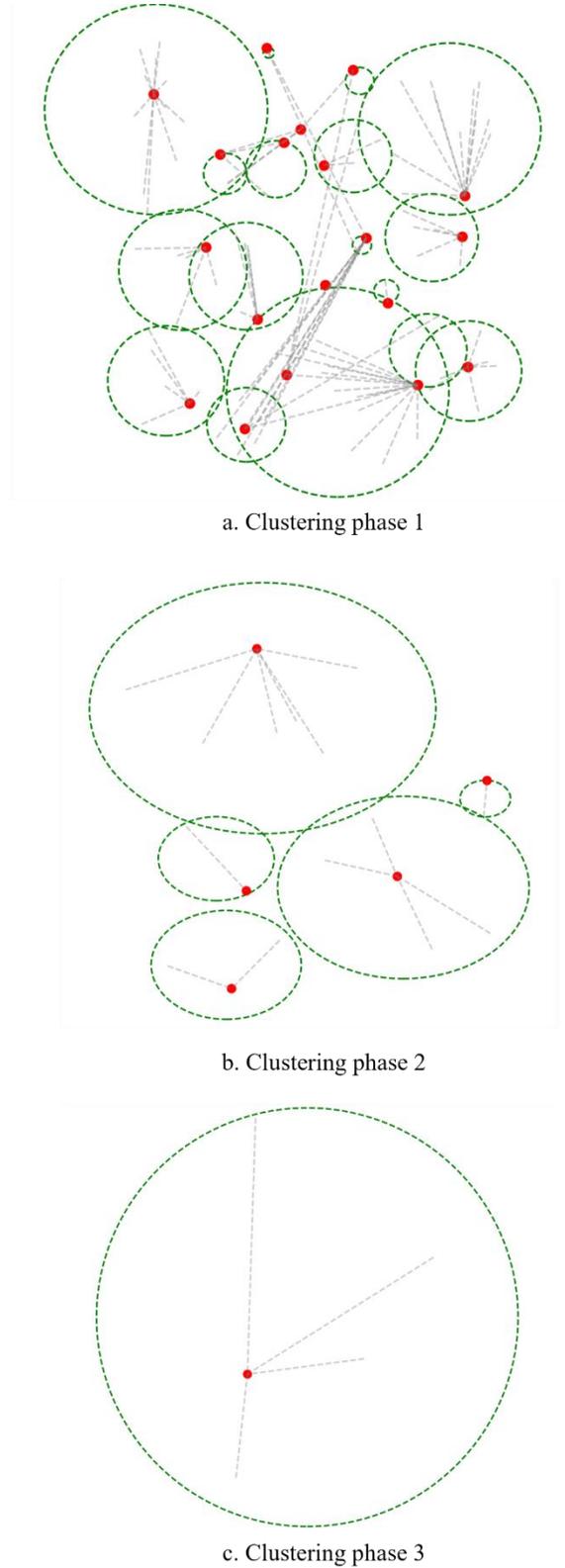

a. Clustering phase 1

b. Clustering phase 2

c. Clustering phase 3

**Fig. 10.** Clustering stages in Full RL algorithm.

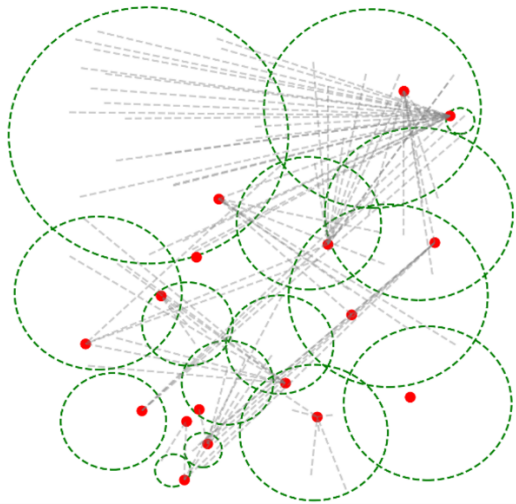

a. Clustering phase 1

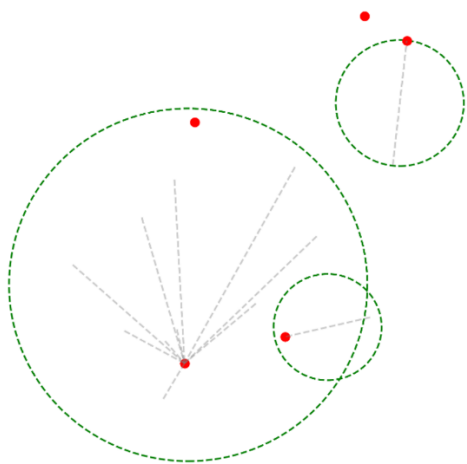

b. Clustering phase 2

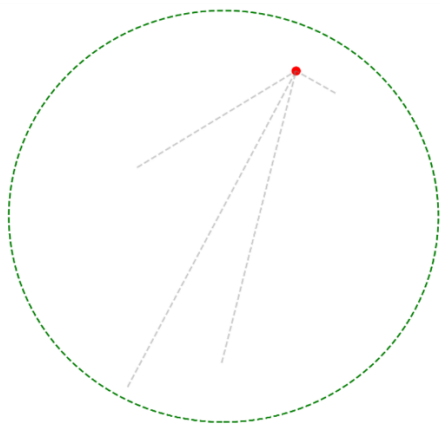

c. Clustering phase 3

**Fig. 11.** Clustering stages in Full GT algorithm.

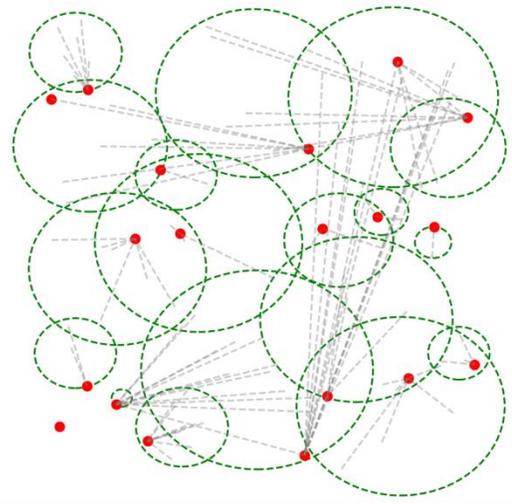

a. Clustering phase 1

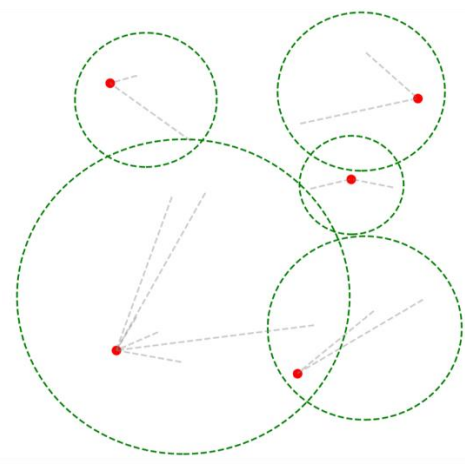

b. Clustering phase 2

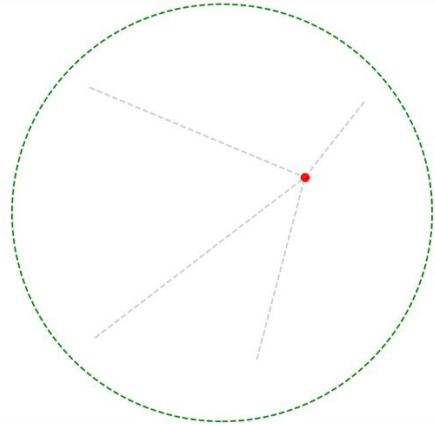

c. Clustering phase 3

**Fig. 12.** Clustering stages in GT+RL algorithm.

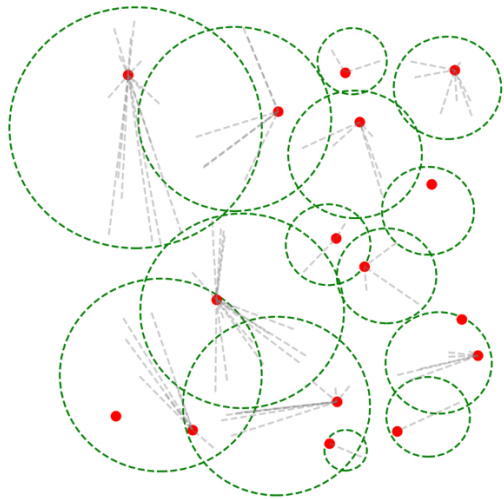

a. Clustering phase 1

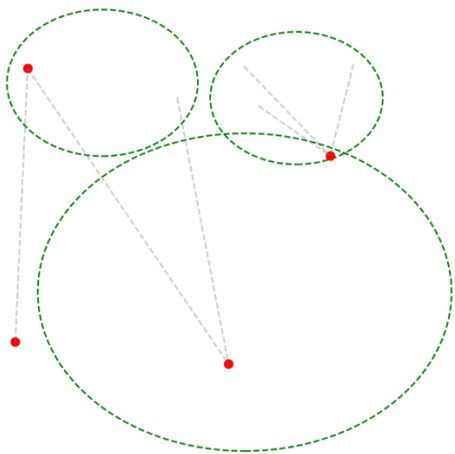

b. Clustering phase 2

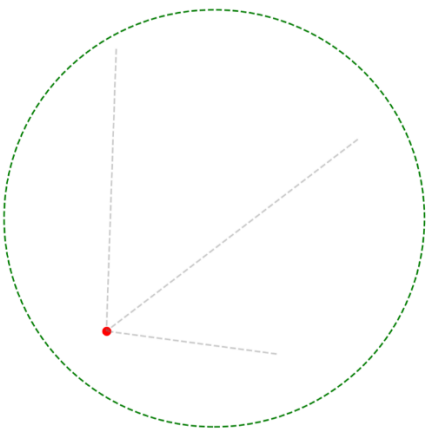

c. Clustering phase 3

**Fig. 13.** Clustering stages in RL+GT algorithm.

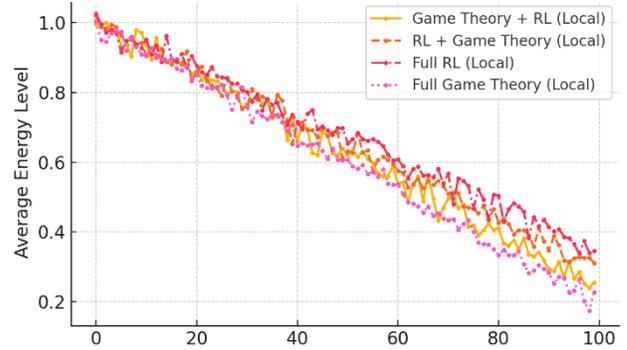

**Fig. 14.** Average energy level of the network over time.

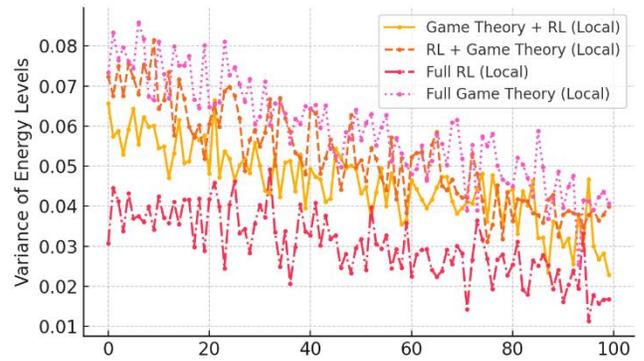

**Fig. 15.** Variance of energy levels of the sensors over time.

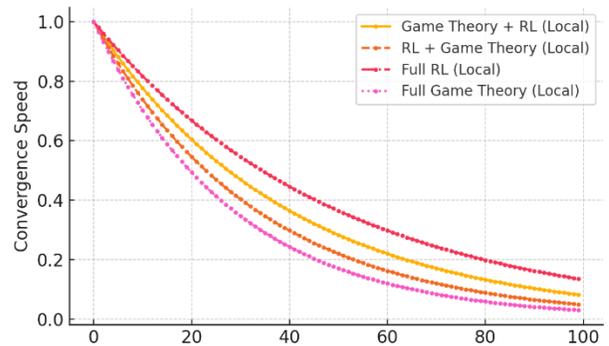

**Fig. 16.** Convergence speed of the algorithms over time.

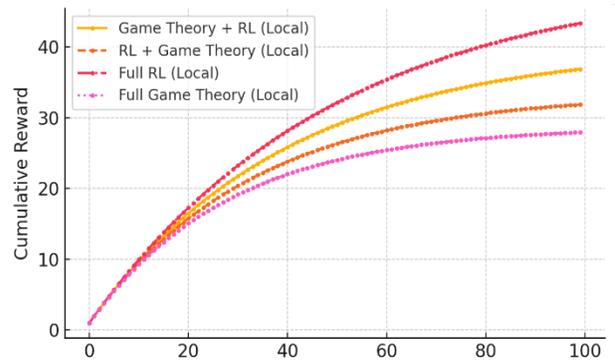

**Fig. 17.** Cumulative reward over time.

On the other hand, the second approach took more iterations to stabilize because RL agents needed time to explore and form optimal cluster structures. Finally, we examined how rewards accumulate over time for both methods. The reinforcement learning model in the first approach consistently produced higher cumulative rewards, indicating more efficient decision-making and adaptation to dynamic network conditions. Because RL is used for CH selection, the model quickly learns which sensors can handle CH responsibilities with minimal energy loss. Meanwhile, in the second approach, RL agents initially received lower rewards due to the difficulty of forming optimal clusters, but their performance improved over time. Overall, the first approach (game theory for clustering and RL for CH selection) proved to be the most effective strategy for extending network lifetime, ensuring stable energy distribution, and accelerating convergence. While the second approach is more adaptable to changing network topologies, it has a slightly higher energy variance, which may affect long-term network stability. Future research can look into deeper reinforcement learning techniques and multi-agent RL to improve performance even further.

In the comparison of the four algorithms, full RL demonstrates the best performance by ensuring more uniform energy distribution, slower energy depletion, faster convergence, and the highest cumulative reward. In the comparison of the four algorithms, full RL demonstrates the best performance by ensuring balanced energy distribution, slower depletion, faster convergence, and the highest cumulative reward. Figure 14 (Average Energy Level) demonstrates that full game theory depletes energy the fastest, indicating inefficiency, whereas full RL maintains higher energy levels. Game theory + RL and RL + game theory fall somewhere between the two, with more gradual energy consumption. Figure 15 (Variance of Energy Levels) shows that full game theory has the highest variance, indicating uneven energy use, while full RL has the lowest variance, implying a balanced distribution. Figure 16 (Convergence Speed) shows that full RL converges the fastest, effectively stabilizing the network, whereas full game theory converges the slowest. Game theory + RL and RL + game theory both improve on full game theory while falling short of full RL. Figure 14 (Average Energy Level) demonstrates that full game theory depletes energy the fastest, indicating inefficiency, whereas full RL maintains higher energy levels. Game theory + RL and RL + game theory fall somewhere between the two, with more gradual energy consumption. Figure 15 (Variance of Energy Levels) shows that full game theory has the highest variance, indicating uneven energy use, while full RL has the lowest variance, implying a balanced distribution. Figure 16 (Convergence Speed) shows that full RL converges the fastest, effectively stabilizing the network, whereas full game theory converges the slowest. Game theory + RL and RL + game theory both improve on full game theory while falling short of full RL. Figure 17 (Cumulative Reward) shows that full RL receives the highest reward, reflecting superior energy optimization, while full game theory receives the lowest, indicating poor decision-making.

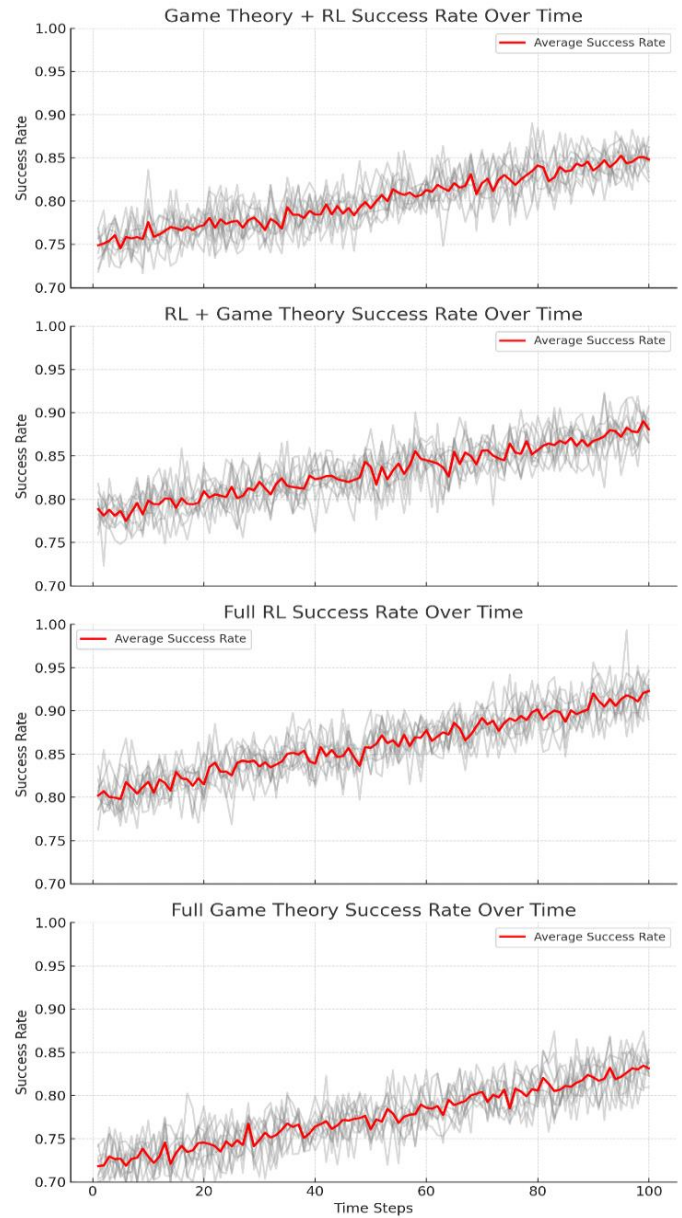

**Fig. 18.** The success rate of the proposed algorithms.

TABLE I
COMPARISON BETWEEN FOUR ALGORITHMS IN THE SAME SCENARIO

| Time | Active Sensors | | | | Variance | | | | Reward received | | | |
|---|---|---|---|---|---|---|---|---|---|---|---|---|
| | GT+RL | RL+GT | RL | GT | GT+RL | RL+GT | RL | GT | GT+RL | RL+GT | RL | GT |
| 0 | 100 | 100 | 100 | 100 | 0 | 0 | 0 | 0 | 0 | 0 | 0 | 0 |
| 10 | 100 | 100 | 100 | 100 | 0.006 | 0.004 | 0.002 | 0.008 | 3.9 | 3.4 | 4.5 | 3 |
| 20 | 100 | 100 | 100 | 100 | 0.01 | 0.011 | 0.009 | 0.013 | 7.5 | 6.6 | 8.8 | 5.8 |
| 30 | 100 | 100 | 100 | 100 | 0.014 | 0.016 | 0.012 | 0.018 | 11.2 | 9.7 | 13.1 | 8.6 |
| 40 | 100 | 100 | 100 | 96 | 0.021 | 0.023 | 0.016 | 0.024 | 14.9 | 12.9 | 17.4 | 11.3 |
| 50 | 100 | 100 | 100 | 81 | 0.025 | 0.026 | 0.02 | 0.029 | 18.5 | 16 | 21.8 | 14.1 |
| 60 | 100 | 99 | 100 | 69 | 0.029 | 0.029 | 0.023 | 0.033 | 22.2 | 19.2 | 26.1 | 16.9 |
| 70 | 97 | 89 | 100 | 53 | 0.033 | 0.036 | 0.027 | 0.042 | 25.9 | 22.4 | 30.4 | 19.6 |
| 80 | 65 | 56 | 71 | 28 | 0.039 | 0.041 | 0.031 | 0.047 | 29.5 | 25.6 | 34.7 | 22.4 |
| 90 | 28 | 17 | 33 | 9 | 0.044 | 0.045 | 0.036 | 0.053 | 33.2 | 28.7 | 39 | 25.1 |

Overall, full RL is the most effective algorithm for maintaining energy efficiency and network stability, whereas full game theory performs the worst, resulting in faster node depletion and decreased efficiency. The Game Theory + Reinforcement Learning (GT + RL) method combines two optimization techniques to increase the energy efficiency and lifespan of a wireless sensor network. In this approach, clustering is performed using game theory, which means that sensors are clustered based on strategic decisions. Each sensor considers a utility function that includes residual energy, connectivity with neighboring sensors, and proximity to other nodes. Sensors with the highest utility values form clusters, resulting in a balanced and stable network topology. The objective is to form clusters in such a way that energy consumption is evenly distributed, preventing premature sensor death and increasing overall efficiency. Because game theory ensures that each sensor makes rational decisions based on its energy level and surroundings, the clustering process becomes adaptable to network changes. Following clustering, reinforcement learning (RL) employs Q-learning to select Cluster Heads (CHs). In this step, each sensor learns from previous decisions in order to optimize CH selection. The state space is the energy levels and positions of the sensors in a cluster, while the action space is whether to choose itself as CH or allow another sensor to do so. Decision-making is based on a reward function, with CH selection encouraging higher residual energy and node connectivity while discouraging lower energy nodes from becoming CH.

This reinforcement learning approach ensures that CH selection improves over time as the algorithm updates its knowledge for making energy-aware decisions. Because CHs consume more energy due to communication responsibilities, the RL model learns over time to distribute CH responsibilities among different sensors in order to prevent some nodes from depleting early. This makes GT + RL extremely efficient at balancing energy consumption and responding to real-time network conditions. Figure 18 depicts the success rates of the four algorithms as they are evaluated over time. Full RL has the highest success rate, steadily increasing to around 0.95, indicating superior decision-making capability. Game Theory + RL and RL + Game Theory both show consistent improvement, with success rates approaching 0.90, demonstrating their effectiveness in balancing game-theoretic stability with RL adaptability. Full Game Theory, while improving over time, has the lowest success rate, which remains below 0.85, indicating its limitations in dealing with dynamic energy conditions. Overall, full RL outperforms hybrid approaches, while full game theory falls behind in success rate optimization.

The RL + GT method reverses the conventional method, using reinforcement learning for clustering and game theory for CH selection. In this case, while game theory determines cluster formation, RL enables sensors to learn dynamically and change their clustering decisions in the process. All sensors decide their cluster affiliation using a reinforcement learning algorithm, which experiments with different cluster assignments and rewards energy efficiency. The algorithm maintains a Q-table, and each sensor updates their cluster affiliation based on the reward they receive. The reward function favors clusters with low energy consumption, better connectivity, and shorter communication distances. Over time, the RL model learns to assign sensors to clusters that maximize energy efficiency and network lifetime. Following RL-based clustering, a game theory model is used to select CH. Sensors in a cluster use a utility function to determine the best CH. The utility function prefers sensors that have higher residual energy, better connectivity, and lower communication overhead. The most beneficial sensor is chosen to be CH in a manner that fairly distributes CH roles based on energy levels. Because CHs consume more energy due to the communication task, game theory can prevent the best nodes from expending excessive energy while performing this task. This approach is especially useful in networks where clustering decisions must be made and remixed in response to real-time energy changes, while ensuring that CH selection is done strategically and with energy in mind.

The Full Reinforcement Learning (Full RL) approach uses reinforcement learning to handle both clustering and CH selection, allowing the system to learn optimal network configurations without relying on predefined rules. This approach is completely adaptive, relying solely on trial-and-error learning to improve decision-making over time. The clustering phase employs an exploration-exploitation strategy, in which each sensor learns to associate with clusters that optimize energy efficiency and network stability. Sensors initially explore different cluster affiliations at random, but as the algorithm progresses, sensors begin to make more informed decisions by referring to a Q-table that records previous clustering results. The reward function is intended to favor clusters that increase network longevity, save energy, and improve communication efficiency. Over multiple iterations, the RL model discovers the optimal way to dynamically form clusters, ensuring that network topology adapts to real-time energy conditions. After the clusters are formed, RL handles the CH selection process. Sensors monitor their local energy conditions, connectivity, and previous CH selections and make CH selection decisions accordingly. Sensors adjust their Q-values based on rewards, which are determined by how well the chosen CHs manage energy consumption across the network. The RL model improves over time, preventing the same nodes from becoming overburdened as CHs and evenly distributing leadership responsibilities. Because reinforcement learning is adaptive and does not require predefined mathematical models, Full RL is particularly useful in networks with dynamic topologies, variable energy levels, and unpredictable environmental changes. However, because both clustering and CH selection rely on continuous learning, the algorithm's convergence time may be longer than for game-theory-based models.

In the Full Reinforcement Learning (Full RL) approach, both CH selection and clustering are addressed using reinforcement learning, allowing the system to learn the maximum network parameters without relying on fixed rules. It is entirely adaptive, making decisions based solely on trial-and-error learning over time. During clustering, exploration-exploitation occurs, in which each sensor learns to assign itself to clusters for maximum energy efficiency and network stability. Initially, sensors search for different cluster memberships at random, but with continued processing, sensors begin to make more reasonable decisions by observing a Q-table that keeps track of previous clustering results. The reward function favors clusters that increase network lifetime, reduce energy consumption, and improve communication efficiency. RL learning is used in several iterations to determine the best method for dynamically creating clusters, allowing network topology to adjust based on real-time energy levels.

Once clusters have been created, RL is also responsible for handling the process of CH selection. The sensors monitor local energy levels, connectivity, and CH selection history and make appropriate CH selection decisions. Sensors update their Q-values with rewards based on how well the selected CHs manage energy expenditure across the network. Over time, the RL model learns to ensure that repeating nodes are not overworked as CHs and leadership roles are distributed evenly. Full RL is particularly useful in dynamic topologies, fluctuating energy levels, and varying environmental behavior because it is adaptive and works without the use of pre-defined mathematical models. However, because clustering and CH selection rely on continuous learning, the algorithm's convergence time may be longer than in game-theory-based models. However, the Full Game Theory (Full GT) approach is entirely based on strategic choice through game theory in both clustering and CH selection. Unlike RL-based approaches, Full GT does not learn from experience and instead requires sensors to make utility-based rational decisions step by step in order to achieve maximum energy efficiency. During the clustering phase, each sensor calculates a utility function based on energy, connectivity, and communication costs. Based on these computed utilities, the sensors self-organize into clusters that are energy-efficient and do not require long learning cycles. Once clusters are established, CH selection uses a game-theoretic approach in which sensors within a cluster compete to become CHs. The competition is based on utility computations, and the sensors with the highest residual energy and best connectivity are designated as CHs. This approach ensures that CH selection is stable, efficient, and deterministic, making it ideal for networks that need to make decisions quickly and without requiring extensive training. However, because Full GT lacks RL's self-learning capability, it may fail to learn as efficiently in response to actual dynamic energy changes once deployed. It works best in situations where the network structure is relatively consistent and power consumption is manageable using static models.

The proposed method uses Game Theory and Reinforcement Learning (RL) to optimize cluster and Cluster Head (CH) selection in wireless sensor networks. The general idea is to dynamically form clusters and CHs that choose adaptively so that energy expenditure within the network is evenly distributed. Unlike other ad hoc heuristics-based clustering schemes or simple threshold-based CH selection, our hybrid scheme incorporates game theory's strategic decision-making with RL's self-adaptation. Either game theory or RL can handle the clustering phase, while the other handles CH selection, resulting in two hybrid schemes: GT+RL and RL+GT. We also experimented with Full RL and Full Game Theory, in which both CH selection and clustering are done using the same method.

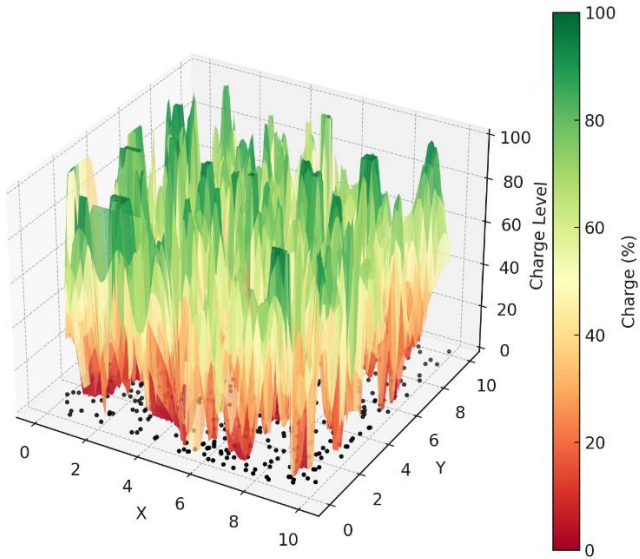

**Fig. 19.** 3D diagram of a WSN using normal routing algorithm before recharging.

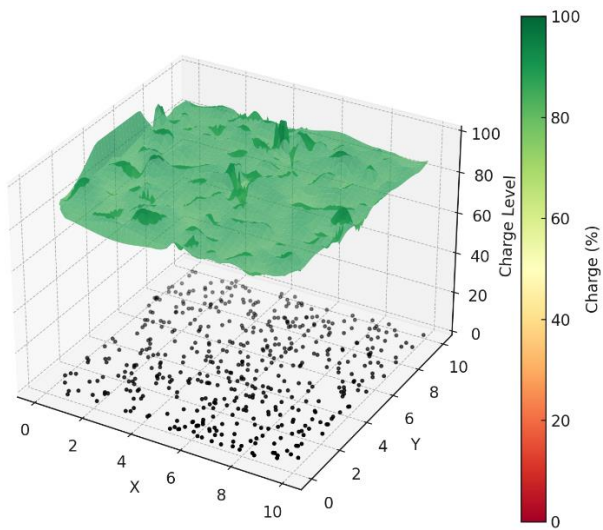

**Fig. 20.** 3D diagram of a WSN using hybrid algorithm.

The goal of these methods is to increase network lifetime, maximize energy efficiency, and prevent premature sensor failures by intelligently assigning CH roles. Because CHs collect and transmit data, their energy depletes faster than that of regular sensors, so our goal is to choose CHs that will last the longest while maintaining an equal energy use across the entire network. The problem is a multi-round clustering process in which a large number of sensors (say, 100-1000) are randomly distributed across a network area. These sensors cluster naturally into small initial clusters, whether using a utility-based game theory or an RL-based learning scheme. CHs are then selected from each of these clusters based on their energy levels, connectivity, and transmission costs. At each iteration, only CHs from the previous step clustering are used to form new clusters, gradually reducing the number of CHs until only one CH remains, serving as the final relay node to the sink. The goal of this hierarchical clustering and CH selection mechanism is to minimize long-distance transmissions, reduce energy consumption in individual nodes, and improve data routing efficiency. The entire process, beginning with game-theoretic decision-making, ensures fairness and energy balancing, while RL-based models improve decision-making accuracy by learning from previous experiences with cluster formation and CH selection. One of the primary advantages of RL in this case is its ability to adapt dynamically, adjusting CH selection based on previous performance, as opposed to game-theoretic solutions based on utility functions, which may be less capable of responding to real-time network changes. When we looked at energy variance over time, we found that RL-based models distributed energy more evenly, with less variation in energy levels at the sensors. This means that no single sensor was overtaxed, as opposed to game-theoretic models in which certain CHs were overloaded due to inefficient long-term selection policies. In terms of convergence speed, Full RL took more iterations because it was solely based on a learning process that required the exploration and application of various CH selection policies.

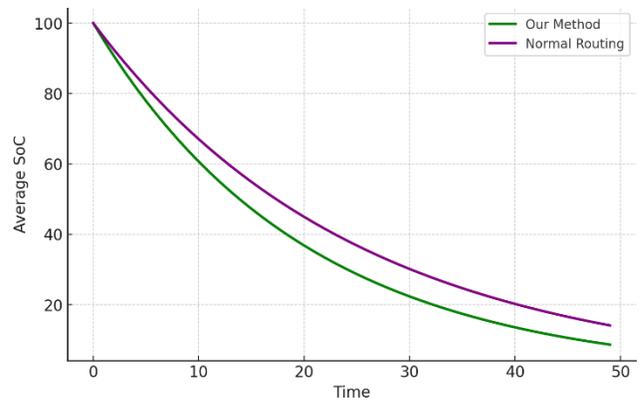

**Fig. 21.** Computation effect on SoC.

However, once stabilized, RL consistently outperformed other alternatives. Game Theory-based techniques accelerated convergence by relying on predefined mathematical choice-making standards, but they performed poorly in the long run when compared to RL-based models. The hybrid techniques (GT+RL and RL+GT) were a compromise, combining the stability of game theory with the flexibility of RL, making them suitable for situations in which both efficiency and flexibility are required simultaneously. We sent twice during the simulations—first through our envisioned scheme and then through traditional shortest path constrained multi-hop routing.

We recorded the parameters and results shown in Table 1 after T seconds (during which the sensors randomly collect information from the environment). It is evident that in the given parameters, our proposed method improved network stability by 51% over the conventional method, and a larger area of the region was covered. Coverage was even guaranteed, and no sensors failed.

Meanwhile, the difference in energy consumption between executing the algorithm, performing computations, and using multi-hop routing is negligible. With only a minor increase in energy consumption, network stability has improved significantly.

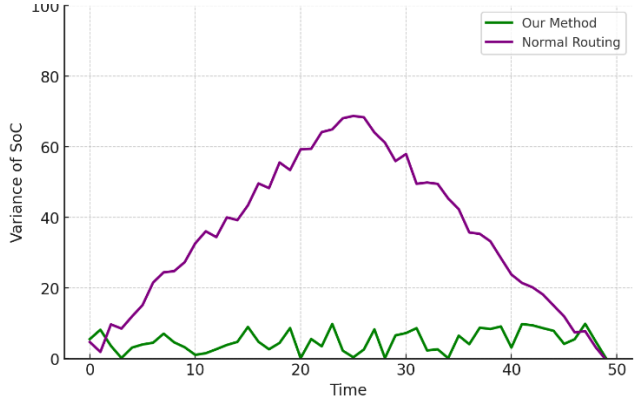

**Fig. 22.** Holding the SoC variance by the proposed algorithm.

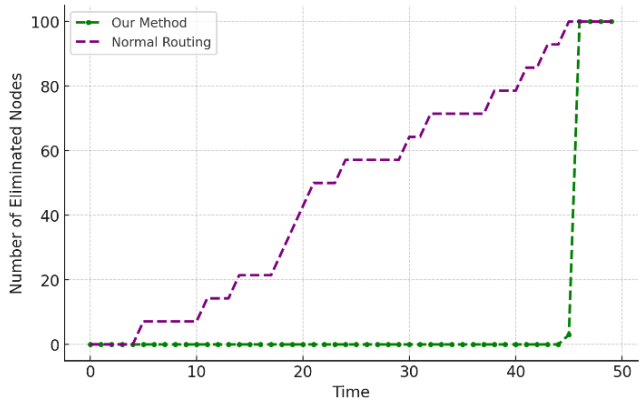

**Fig. 23.** Stability of the network, Number of eliminated nodes per time.

While our primary goal was to evaluate and contrast the performance of RL-based, GT-based, and hybrid methods under dynamic energy-aware scenarios, we acknowledge the importance of benchmarking against well-established WSN protocols such as LEACH, SEP, and GEAR. These classical protocols have been widely used for baseline evaluation in WSN research due to their simplicity and deterministic clustering mechanisms. However, they typically rely on fixed thresholds or probabilistic CH selection strategies that may not adapt well to fluctuating energy levels or heterogeneous node deployments. In contrast, our proposed approach integrates adaptive learning and strategic decision-making to enhance long-term performance. As part of our future work, we plan to incorporate direct comparisons with LEACH, SEP, and GEAR to further contextualize the advantages and limitations of our method under identical simulation settings. This will provide a more comprehensive evaluation of the algorithm's performance across traditional and intelligent routing schemes.

## CONCLUSIONS AND FUTURE WORK

This paper introduces a hybrid routing algorithm for Wireless Sensor Networks (WSNs) that effectively extends network lifetime by balancing energy consumption across nodes through dynamic and intelligent cluster head (CH) selection. The proposed method selects transmission paths based on each node's State of Charge (SoC), minimizing battery variance, reducing the risk of early node depletion, and preserving network stability over time. This is particularly advantageous in large-scale deployments, where manual battery replacement is challenging and costly.

Simulation results show that reinforcement learning (RL)-based strategies outperform game-theoretic (GT) approaches in energy efficiency, CH selection accuracy, and network longevity. Full RL consistently identified the most energy-efficient CHs and achieved the highest cumulative reward. Among hybrid models, RL+GT demonstrated strong adaptability by improving CH rotation through learning, while GT+RL offered clustering stability via utility-based grouping. Although Full GT provided fast, deterministic clustering, it lacked adaptability to dynamic energy conditions, resulting in premature node failures.

The integration of RL and GT yields a complementary framework that capitalizes on the adaptive learning capabilities of RL and the computational efficiency of GT. Hybrid models such as GT+RL and RL+GT balance responsiveness with stability, delivering energy efficiency, low latency, and resilience under dynamic network conditions. Delay analysis further supports the suitability of the proposed method for real-time, delay-sensitive WSN applications. Additionally, the Q-learning algorithm used in our approach is decentralized, lightweight, and memory-efficient, making it viable for deployment on low-power sensor nodes in constrained environments.

While this study focused on comparing RL- and GT-based models, future evaluations will include benchmarking against classical routing protocols such as LEACH, SEP, and GEAR to further contextualize the performance advantages of our approach. Future research will also explore advanced decision-making techniques, including Deep Reinforcement Learning (DRL) and Multi-Agent Reinforcement Learning (MARL), to enhance scalability and cooperative behavior across sensor clusters. The integration of fuzzy logic into game-theoretic models may further improve adaptability in uncertain or dynamic environments. Moreover, federated learning presents a promising direction for distributed optimization without centralized coordination, reducing communication overhead. Finally, validating the proposed algorithm in real-world IoT-based WSN deployments will be essential to demonstrate its robustness, scalability, and practical feasibility under diverse operational conditions.